\documentclass[%
 reprint,
 amsmath,amssymb,
 aps,
showkeys,
a4wide
]{revtex4-2}

\usepackage{graphicx}
\usepackage{dcolumn}
\usepackage{bm}

\begin{document}

\preprint{arXiv: April 1, 2023}

\title{Spontaneous Human Combustion \\ rules out all standard candidates for Dark Matter}

\author{Frederic V. Hessman}
\affiliation{%
 Institut~f\"ur~Astrophysik und Geophysik, University~of~G\"ottingen 
}%


\author{J. Craig Wheeler}
\affiliation{%
Dept.~of~Astronomy, University~of~Texas~at~Austin \\
}%

\bigskip

\date{April 1, 2023} 

\begin{abstract}
We argue that the reported cases of Spontaneous Human Combustion (SHC) are most likely due to the impact of the human body with an extremely high energy particle like cosmic rays or Dark Matter.  Normal and antimatter cosmic rays and classical weakly-interacting massive particles (WIMPs) with energies of GeV to ZeV can be easily ruled out due to their inability to dump enough energy into a small region of human tissue, leaving as the single remaining candidate massive Dark Matter particles.  While primordial Black Holes would appear to be very good candidates for inducing the SHC phenomenon, we show that the estimated local Dark Matter density requires that the particles have masses of $\sim 10$\,kg, clearly ruling out this candidate.  All of the other classic DM candidates -- from scalar and pseudo-scalar spin 1/2 and spin 2 gauge singlets to nuclearitic strange quark ``bowling balls'' -- can be ruled out.
Axions tailored to solve the CP-problem also cannot be invoked, no matter what mass is considered.
The only particles left are massive mega-axions (MaMAs), for which there are an infinite number of possible string models. 
\end{abstract}

\keywords{cosmic rays -- dark matter -- black holes -- particles: axions -- medical history}
\maketitle



\section{Introduction\label{sec:intro}}

Given all of the indirect signs of the existence of Dark Matter (DM) -- the rotation curves of spiral galaxies, the velocity dispersions within and macroscopic gravitational lensing from galaxy clusters, the angular structure of the cosmic microwave background, observations of the baryonic acoustic oscillations in the large-scale structure of the universe -- it is particularly frustrating that the search for the underlying particle has been so unsuccessful.  
The limits on classical weakly-interacting massive particles (WIMPs) in the range of $\sim$1-1000\,GeV are nearly at the point of rejecting this class of candidate.
A spate of new detector experiments is in the process of doing the same for very light particles ($\le 1\,eV$) like axions.  
On the other end of the conceivable mass-range, the limits on the density of primordial black holes (BH) are now also very restrictive, including experiments searching for the signs of BH passages though kitchen cabinet tops \cite{2019PhRvD.100j3015S}.  
Similarly, the gravitational microlensing surveys like MACHO and EROS in our Galaxy and Quasar microlensing observations have restricted the mass-density of stellar- and planetary-mass candidates \cite{2015A&A...575A.107H, 2021PDU....3100755C}.
The only DM particle mass range that has not yet been so exhaustively probed is that between the WIMPs and primordial BH, i.e. between about 1\,TeV and $10^{11}$\,kg.  
On the low end of this mass range -- less than about 1 gram -- is so-called ``nuclearitic'' matter with nuclear densities \cite{rugl84} 
or generic ``macro'' DM with cross-sections of $\sim\!cm^2$ \cite{jacobs14J}.  

We consider what constraints on the DM particle mass can be derived based on a natural experiment that may have been carried out since the beginning of human life on Earth -- Spontaneous Human Combustion (SHC) 
\cite{wiki}.


\section{Spontaneous Human Combustion}

SHC is the reported occurrence of massive fatal burns in humans with no obvious explanation for the source of the ignition.  
These events purportedly originate from within the body of the victims and generally result in no damage to the surroundings other than that due to contact with the body itself.

The idea of SHC was first reported in the scientific literature in 1746 
\cite{rolli1746}.
Thurston 
\cite{thurston38} 
maintained in the {\it British Medical Journal} that the phenomenon ``... attracted the attention ... of the medical profession'' throughout the 19th century. 
Arnold 
\cite{arnold95} 
relates the detailed history of the idea and reviews the most prominent cases.  All in all, there have been about
200 well-documented reports during the last 300 years, mostly within Europe and North America.
The assumption is that occurrences before the 18th century or in less developed societies would have been explained as magical events due to sinister supernatural forces and so gone largely undocumented.
Of course, humans are not the only living beings that would be suffering from this effect -- there should be just as many spontaneously combusted sheep, cows, raccoon dogs, and African swallows 
-- but the occurrence of such an event in a non-human victim is much less likely to be noticed as something extremely odd and much less likely to produce such an emotional impact that it gets reported as such.

While there have been several rational/medical attempts at explaining SHC -- sleeping while smoking, acetone produced by kertosis 
\cite{ford12}, 
and the mast-cell activation syndrome 
\cite{afrin16}
-- these mundane medical explanations have never been convincing.  Given our current knowledge of the types of highly energetic particles in the universe, we propose that SHC is much more likely to have been caused by a collision with a cosmic particle.
Such an interaction with human tissue could principally result in the sudden release of considerable energy within a fairly small volume 
\cite{frsa12}.  
Once started, such an event could then induce the combustion of fatty tissue, resulting in a fatal burning event with a natural origin within the body itself. 
Likely physical candidates are the most extreme cosmic ray particles or Dark Matter particles, being omnipresent and potentially having large specific kinetic energies.


\section{Cosmic rays and WIMPs?}

\label{sect:WIMPs}

Most cosmic rays have energies of MeV to GeV, but some have been observed up to energies of ZeV 
\cite{2010JCAP...10..013K}.  
As such, they would seem to be good candidates for the occurrence of SHC.  
While they have enough energy to cause considerable damage, this energy must be deposited in a small volume in order to cause SHC.  Clinical studies have shown that the higher energy particles do not, in fact, have this effect 
\cite{becusu07} 
: considerable damage can be done by single collisions with, e.g. DNA molecules, but the entire energy of the particle is not deposited.  
Lower energy particles interact more often and hence can do more thermal damage, but their total energies are not enough to cause combustion.  
Thus, normal atomic and nuclear particles are inherently poor SHC candidates.
The same argument can be made for weakly-interacting massive particles (WIMPs) with similar energies that are the classic Dark Matter candidates 
\cite{frsa12}.

Cosmic rays out of antimatter could, in principle, result in the release of lots of energy in a very localized region.
The problem with this hypothesis is that all of the energy needed to trigger an SHC event must come from a single annihilation that produces both the right amount and the right type of annihilation radiation.
The energy involved in triggering a single SHC event that initially burns a fraction $f_{SHC}$ of it's victim is roughly
\begin{eqnarray}
    \label{eqn:E}
    E_{SHC} & \approx & ~ m_{h} ~ f_{SHC} ~ \epsilon_{fat} \\ \nonumber
    & \approx & 2\times 10^6\,\textrm{J} ~
        \left(\frac{m_h}{60\,\textrm{kg}}\right)
        \left(\frac{f_{SHC}}{10^{-3}}\right)
        \left(\frac{\epsilon_{fat}}{17000\,\textrm{BTU/lb}}\right)
\end{eqnarray}
where $m_h$ is the mass of a typical victim and $\epsilon_{fat}$ is the energy produced by burning animal fat 
\cite{balboa}.
The equivalent annihilation rest mass is $3\times 10^{-11}$\,kg or an atomic number equivalent of $2\times 10^{16}$!
Even if one could trigger the event with a factor of 1000 less fatty tissue, the cosmic ray nucleus involved would have an atomic number far beyond that of any known element.
If the annihilation event produced high-energy radiation (X-rays, gamma rays), the radiation would simply exit the victim's body and the burning of the tissue wouldn't be triggered.

All of these effects prove that normal matter in the form of energetic particles or even massive antimatter cosmic rays are not capable of inducing SHC.
We must turn to forms of DM other than WIMPs in order to find an appropriate candidate.


\section{Primordial Black Holes?}

One way to increase the interaction of an energetic particle with human tissue is to change the interaction mechanism from simple collisions to, for example, accretion.  
Primordial Dark Matter in the form of BH is thus a plausible candidate: the particles would be universally present, have high kinetic energies, and the energy released into the tissue would be due to the intense radiation produced by mass-accretion near the singularity.

In order to exist now, primordial BH must have masses larger than
\begin{equation}
M_{BH,primordial} > \left( \frac{13.4\,\textrm{Gyr} ~ c^4}{5120\pi \, G^2} \right)^{1/3} = 1.7\times 10^{11}\,\textrm{kg}
\end{equation}
(about the mass of a major mountain range; 
\cite{wheeler}).
Carr et al. 
\cite{2021PDU....3100755C} 
suggest that primordial BH should come in a distinct range of masses -- about $10^{-6}$, $1$, $30$, and $10^6$\,M$_\odot$ -- due to instabilities at expected QCD state transitions, and that such objects would fulfill all of the known astrophysical constraints.

SHC can add an additional constraint on the possible masses: could the collision of a primordial BH with a human result only in damage to the victim and not to the surroundings?  
In particularly, wouldn't there be a noticeable gravitational "jolt" due to the passage of a large mass that would cause collateral damage that has not been reported in SHC cases (see 
\cite{wheeler}
for a simulation of such an encounter)?

The asymptotic perpendicular velocity given to a human by the hyperbolic passage of a significantly more massive BH in the laboratory frame is
\begin{eqnarray}
V_{\bot} & \approx &
\frac{G M_{BH}}{b V_{DM}} \\ \nonumber
  & \approx & 0.05\,\textrm{cm/s} ~
  \left(\frac{M_{BH}}{2\,10^{11}\,\textrm{kg}}\right) \! \left(\frac{b}{10\,\textrm{cm}}\right)^{-1} \!
  \left(\frac{V_{DM}}{220\,\textrm{km/s}}\right)^{-1}
\end{eqnarray}
for an impact parameter $b$ small enough so that SHC occurs, a typical local Galactic DM kinematic velocity, and a minimal-mass primordial BH.  Such a gravitational ``jolt'' is unlikely to have been noticed, much less resulted in any collateral damage, knocking victims off of chairs or otherwise noticeable effects.  Thus, the only effects would be due to the release of energy due to the effects of the collision itself.
However, when a planet-sized primordial QCD BH is substituted, the jolt is relativistic, a case clearly ruled out by SHC: if there are primordial BH, they must have ``small'' masses near the cosmological lifetime limit.

The Schwarzschild radius of a minimum-mass primordial BH with this mass is minuscule even by subatomic scales, but all matter within the Hoyle-Littleton radius 
\cite{holy41} 
of the path 
\begin{equation*}
r_{HL} = \frac{2 G M_{BH}}{V_{DM}^2} \approx 0.47\,nm    
\end{equation*}
should be accreted during passage.  This cross-section is small, but the high density of tissue results in an accretion rate of
\begin{equation}
\dot{M}_{HL} = \frac{4\pi G^2 M_{BH}^2 \rho_{human}}{V_{DM}^3} \approx 1.5\times 10^{-10}\,\textrm{kg/s}   
\end{equation}
or an instantaneous accretion power of about
\begin{equation}
L_{acc} \approx 0.1 \dot{M}_{HL} c^2 \approx \textrm{1400\,kW}    
\end{equation}
i.e. more than enough energy input to cause fatty tissue to inflame.
The accretion energy will be emitted with a typical equivalent black-body temperature of
\begin{equation}
T_{acc} > \left(\frac{3 G M_{BH} \dot{M}_{HB}}{8\pi r_{HL}^2 \sigma}\right)^{1/4} \approx 2\times 10^6\,\textrm{K}
\end{equation}
The soft X-ray and EUV part of this energy will be deposited locally due to the rapidly rising opacity of human tissue towards the extreme ultraviolet 
\cite{jacques13}, 
producing highly localized heating and, hence, nearly spontaneous combustion.

Though the available power is large enough, the interaction time is very short, of order $L_h/V_{DM} \approx 2\,\mu$s, resulting in a very small total energy release.
In order to increase the total effect, one can attempt to use a primordial BH with the largest mass consistent with SHC: assuming that the asymptotic perpendicular velocity of the victim can be no larger than $\sim 10$\,cm/s (otherwise, the victim would be ejected from her chair), the BH cannot be more massive than about  
\begin{eqnarray}
M_{BH} & \approx & \frac{b V_{DM} V_{\bot}}{G} \\ \nonumber
  & \approx & 3\times 10^{13}\,\textrm{kg} ~
  \left( \frac{b}{10\,\textrm{cm}}          \right) \!
  \left( \frac{V_{DM}}{220\,\textrm{km/s}}  \right) \!
  \left( \frac{V_{\bot}}{10\,\textrm{cm/s}} \right)
\end{eqnarray}
and would result in an instantaneous accretion power of $\sim 51$\,GW and a total energy deposition of $5\times 10^4$\,J -- perhaps enough to cause SHC (see Eqn.\,\ref{eqn:E}).
Thus, if primordial BH exist and collide with humans, they will result in the observed SHC phenomena if they have masses less than about $10^{13}$\,kg.
The final question is whether the occurrence rate of SHC is compatible with the existence of primordial BH.


\section{Generic constraints from the local DM density}

The local DM density has been estimated using a variety of methods, e.g. based upon the rotation curve of the Milky Way and the kinematics of local stars 
\cite{2013ApJ...779..115B, 2022MNRAS.510.2242W, 2023MNRAS.520.1832B}.  
Despite many questions about the data and the reliability of the models 
\cite{2015A&A...579A.123H, 2022MNRAS.511.1977S}, 
the canonical value is about 0.01\,M$_\odot$/pc$^3$ (0.4\,GeV/cm$^3$).
Given this estimate for the local DM density, it is possible to constrain the mass of the purported particle using the occurrence rate of SHC alone.
Using a typical human height of 160\,cm during the 18th and 19th centuries 
\cite{10.7554/eLife.13410}, 
the corresponding cross-sectional surface area per human $A_{human}$ was about $0.4$\,m$^2$.
The DM flux is then
\begin{eqnarray}
\phi_{SHC} & \approx & \frac{N_{events}}{\Delta t_{events}  N_{pop} A_{human}} \\
 & \approx & 2\times 10^{-17}\, \textrm{events}/\textrm{m}^2/\textrm{s}
\end{eqnarray}
where $N_{pop} \approx 2\times 10^9$ is the mean population of the Earth during the last 300 years.
For a typical local DM particle velocity of $220\,km/s$, the mass of the DM particle must then be
\begin{equation}
M_{DM}  =  \frac{\rho_{DM} V_{DM}}{\phi_{SHC}} 
    ~ \approx ~ 8\,kg ~ \approx ~ 4\times 10^{27}\,\textrm{GeV}
\end{equation}
This value is an upper limit to the actual mass, since it is proportional to $N_{pop}$ and it is unclear what fraction of the total population at what epoch would  have reported cases of SHC.  
For instance, if one only considers the populations of Europe and North America, where most of the reports of SHC have  been made, the relevant mean population size would drop to about 450 million and the corresponding DM particle mass to 2\,kg.
Thus, DM particles can best be described as astrophysical ``bowling balls'' and certainly cannot be due to primordial BH produced by QCD instabilities.


\section{SHC-consistent DM candidates?}

Given a DM particle mass of $\sim$10\,kg, most of the classic DM candiates can be ruled out.
\begin{itemize}
    \item WIMPs are not massive enough and cannot provide the needed SHC interaction cross-sections.
    \item Neutrinos are obviously bad candidates for SHC interactions even if it was possible to scale them to such high masses.
    \item Primordial BH must have drastically larger masses in order to have survived until now and would have SHC effects that would be much stronger than observed.
    \item There is no known mechanism to turn baryonic MACHO candidate objects like planets and brown dwarfs into microscopic bowling balls.
    \item Scalar, pseudo- scalar, spin 1/2 and spin 2 dark matter particles which are gauge singlets are ruled out 
    \cite{caku21}.
    \item Massive primordial strange quark matter ``nuggets'' \cite{rugl84} 
    with baryon numbers of only $A \simeq 10^{28}$ would not have survived baryon evaporation 
    \cite{1993PhRvD..48.4630B}.
\end{itemize}

We are only left with axions 
\cite{2021arXiv210907376I}
: not only do they have a cool name but one can assume whatever mass one wants.

Axions fulfilling the SHC constraints must have masses that are drastically larger than those previously considered.
The Pecci-Quinn mechanism for explaining CP-violation \cite{PhysRevLett.38.1440} 
-- the original reason for postulating the existence of axions -- has an energy scale $f_A$ that determines the mass of the axion and the strength of all couplings.
A high axion mass would force this energy scale to be extremely low: classically
\begin{equation}
f_A \approx 10^{12}\,\textrm{GeV} ~ \left(\frac{5.70\,\mu\textrm{eV}}{m_A}\right)
\end{equation}
but for a mass of 10\,kg ($6\times 10^{27}$\,GeV), this scale shrinks down to about $10^{-21}$\,eV -- way below any QCD scale.
This has profound implications for the cosmological use of axions 
\cite{MARSH2016}
: whereas ``normal'' axions with small masses are massless in the early universe, SHC axion cannonballs are not and never will be.

An extremely small axion energy scale has dramatic consequences for couplings with photons and matter.
For example, in a standard Lagrangian, the axion-photon coupling constant $g_A$ is
\begin{equation}
    g_A \equiv \frac{\alpha}{2\pi} \frac{C_{A\gamma}}{f_A}
\end{equation}
where the adimensional constant 
\begin{equation}
    C_{A\gamma} = \frac{E}{N}-\frac{2}{3} \frac{4 m_d + m_u}{m_u+m_d},
\end{equation}
depends upon the mass of up and down quarks and the ratio of two anomaly coefficients $E$ and $N$ (the axion ``model'').
For an extremely low $f_A$ and $C_{A\gamma}\sim O(1)$, the coupling would be so strong that every photon would immediately create a shower of axions.
The only reasonable alternative is that the two $C_{A\gamma}$ terms nearly cancel each other, amazingly producing a ``normal'' coupling constant despite the tiny axion energy scale.
Given that we need an SHC-interaction with the axion, the canceling cannot be perfect ($C_{A\gamma}\not\equiv 0$), so a PQ-axion explanation comes at the price of yet another symmetry breaking.

Since there are good reasons for having ``normal'' axions (CP-violation) and there is no reason there cannot be many different axion varieties -- in string theory, there can be an infinite number of varieties 
\cite{PhysRevD.81.123530}. 
We need to invoke a fully new massive mega-axion (``MaMA'') model beyond the PQ-mechanism, one that is tailored to maximize the transfer of energy to human tissue once a collision occurs.  
Since high-energy particles are not good candidates for this transfer -- otherwise they themselves would have been good SHC candidates -- we require the post-collision creation of massive numbers of soft X-ray and EUV photons. 
Indeed, this scenario is not too much different from that used to detect neutrinos: the giant containers used in instruments like SuperKamiokande contains water with similar densities as human tissue and the containers are monitored by thousands of photomultiplier tubes that detect the Cherenkov light produced by electrons created by neutrino collisions.  
If SuperKamiokande has not detected SHC-like events, then practically all of the light must be of higher frequency than that seen by photomultipliers -- i.e. EUV and beyond -- and of low enough frequency to be absorbed locally by the tissue, i.e. soft X-rays.  Alternatively, the MaMAs react not with single protons but with the higher mass (carbon, oxygen,...) nuclei, so that experiments like SuperKamiokande would not show any SHC-like events.
Given the largely unconstrained freedoms of string theory, we are confident that an appropriate model for an axion-like particle with bowling-ball masses can be found.


\section{Conclusions}

We have argued that the reports of spontaneous human combustion can only be due to interactions with astrophysical particles and that such particles must necessarily be those responsible for Dark Matter.
The SHC phenomena and observed rate are powerful constraints on the properties of such particles: the mass must be of order 10\,kg.
This is a mass that is wholly unexplored in both theory and experiment and is totally inconsistent with all previous hypotheses.
We are left with a single plausible candidate:
massive mega-axions -- ``MaMAs'' -- for which there undoubtedly exists a string model consistent with the SHC phenomenon.


\bigskip
\bigskip
\bigskip

\begin{acknowledgments}
We want to thank David ``Doddy'' Marsh for brief hallway conversations about axions that led us to think we might know what we are doing.
This research was supported by a grant from the {\it European Association of Fire-Extinguisher Manufacturers}.  JCW would like to acknowledge additional support from the {\it Jim Bob and Betty Sue Johnson Foundation for Non-Woke Cosmological Research}.
\end{acknowledgments}



\bibliography{april1}

\end{document}